\input psfig.sty
\documentstyle{aipproc}

\begin{document}
\title{Is the Fuzziness of GRB970228 constant?}

\author{Patrizia A. Caraveo$^*$, Roberto Mignani$^{\dagger}$ 
and Giovanni~F.~Bignami$^{+,*}$}
\address{$^*$Istituto di Fisica Cosmica del CNR, Milano, ITALY\\
$^{\dagger}$MPE, Garching, GERMANY\\
$^+$Agenzia Spaziale Italiana, Roma, ITALY}


\maketitle

\begin{abstract}
In view of the data gathered in September 1997, we review the flux  values collected so far
for the "fuzziness" seen in the optical counterpart  of GRB970228. Comparison between the
ground based data collected in  March and the data of September 1997 suggests a  fading of the
fuzz.   Given the diversity of the data in hand, the  magnitude of the effect and its
significance are not easy to quantify. Only new images, both from the ground and with the
Space Telescope,   directly comparable to the old ones could settle this problem.
\end{abstract}

\section*{Introduction}
After the SAX positioning of GRB970228 (Costa et al. 1997), and the  discovery of an optical
transient in the refined error box (van Paradijs  et al, 1997), the optical counterpart of 
GRB970228 has been observed  many times both with ground based instruments and with the
Hubble  Space Telescope. Several days after the event, an extended optical emission was
detected  where the Optical Transient (OT) had been seen in the discovery image, taken  ~21 h
after the event (van Paradijs et al, 1997). Since then, the  magnitude of such an extended
emission has been measured many  times, by several observers, using different instrumental
set-ups.  In this paper we review and compare the measurements gathered so far  to investigate
if the flux values recently measured by STIS on HST  (Fruchter et al, 1997) and by the 5m
Palomar telescope (Djorgovski et  al., 1997) are consistent with the ground based ones
obtained at early  epochs. 

\begin{figure}[b!] 
\centerline{
\psfig{file=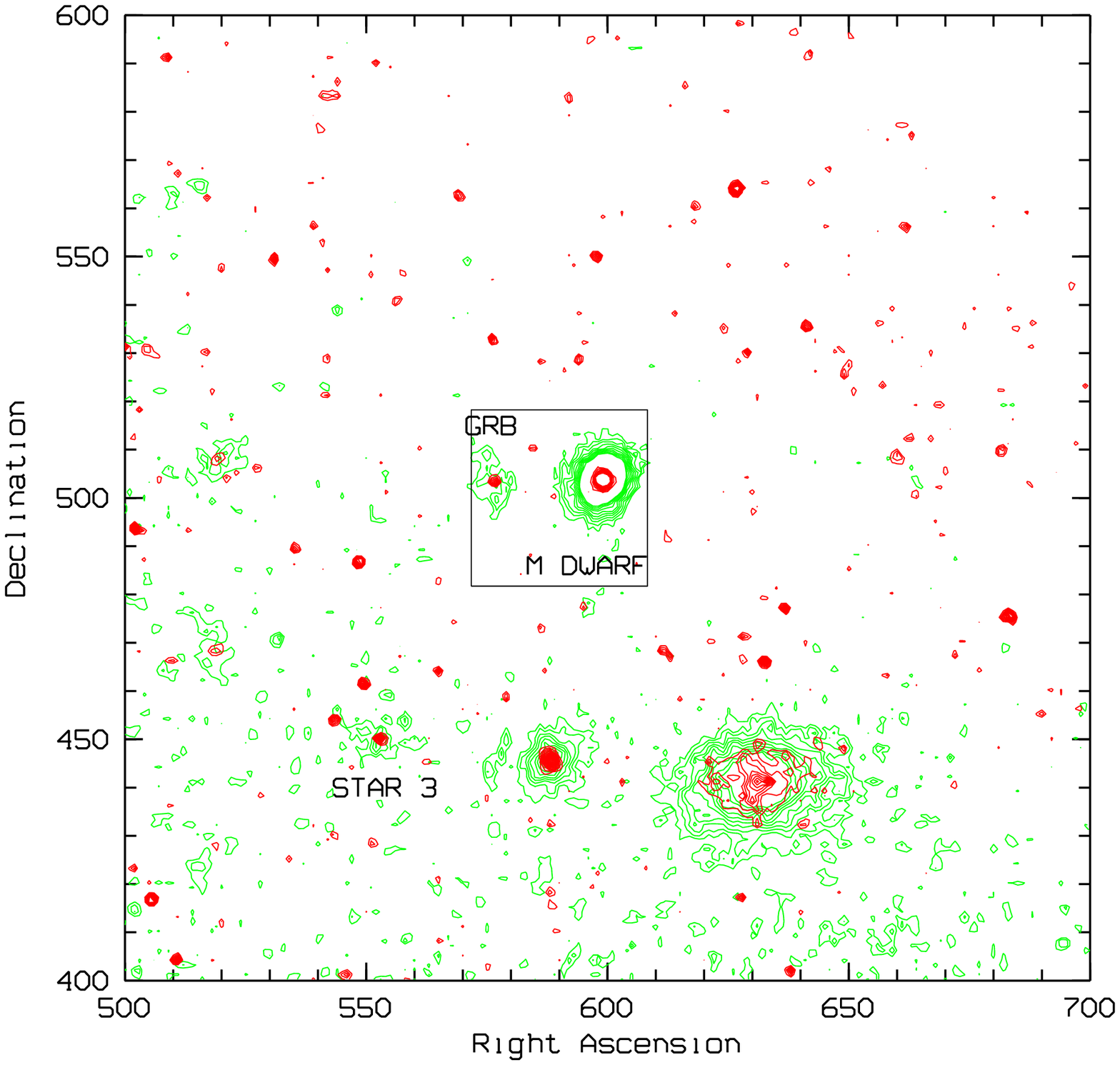,height=6cm,angle=270,clip=}
\psfig{file=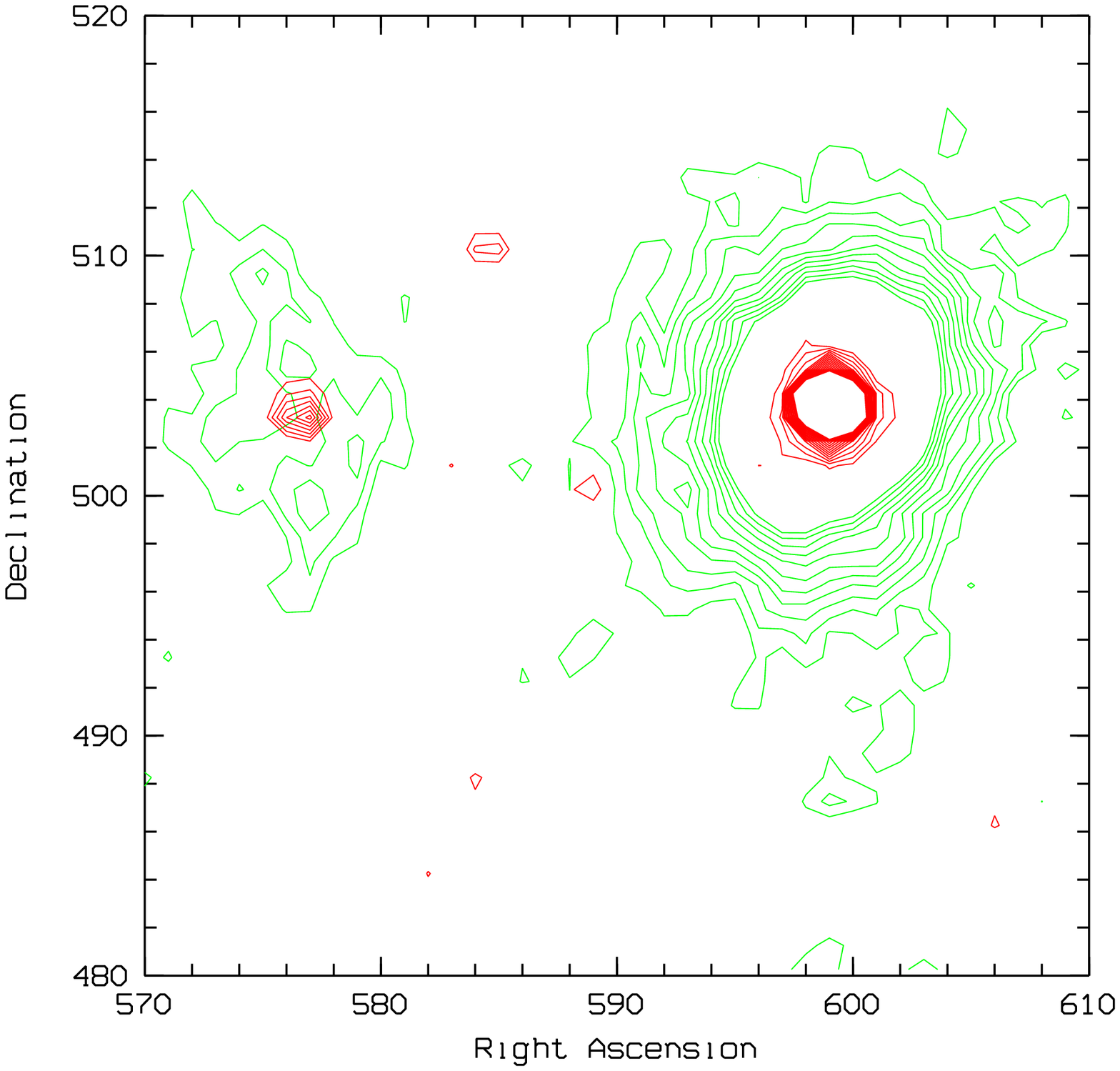,height=6cm,angle=270,clip=}
}

\caption{Superposition of the March HST/PC frame onto
the NTT/SUSI one (actually, only a 26$\times$26 arcsec  area is shown).  
North to the top and East to the left. Axis units are SUSI pixels  (0.13 arcsec).
The GRB counterpart as well as the nearby M dwarf and star \#3 (see text) are labelled.  A
zoom of the central square is shown in the right panel. The GRB position measured by the
HST/PC falls exactly at the center of the  nebulosity observed by the NTT. }
\label{}
\end{figure}

\section*{The Data}
Table 1 summarizes the data collected so far, both for the OT  integrated magnitude (ground
measurements) and for the contribution of  the two components : point source and extended
emission (HST data).  The first claim for an extended object, using an 1 hour NTT exposure 
taken on March $13^{th}$, gave $m_{R}=23.8 \pm 0.2$ (van Paradijs et al 1997).  To  this, 
one should
add the Keck measurement ($m_{R}=24.0 \pm 0.2$, Metzger et  al, 1997a) obtained on 
March $6^{th}$ but
announced after the discovery of the  optical transient (Groot et al. 1997a). 

 \begin{table}
 \centerline{\psfig{figure=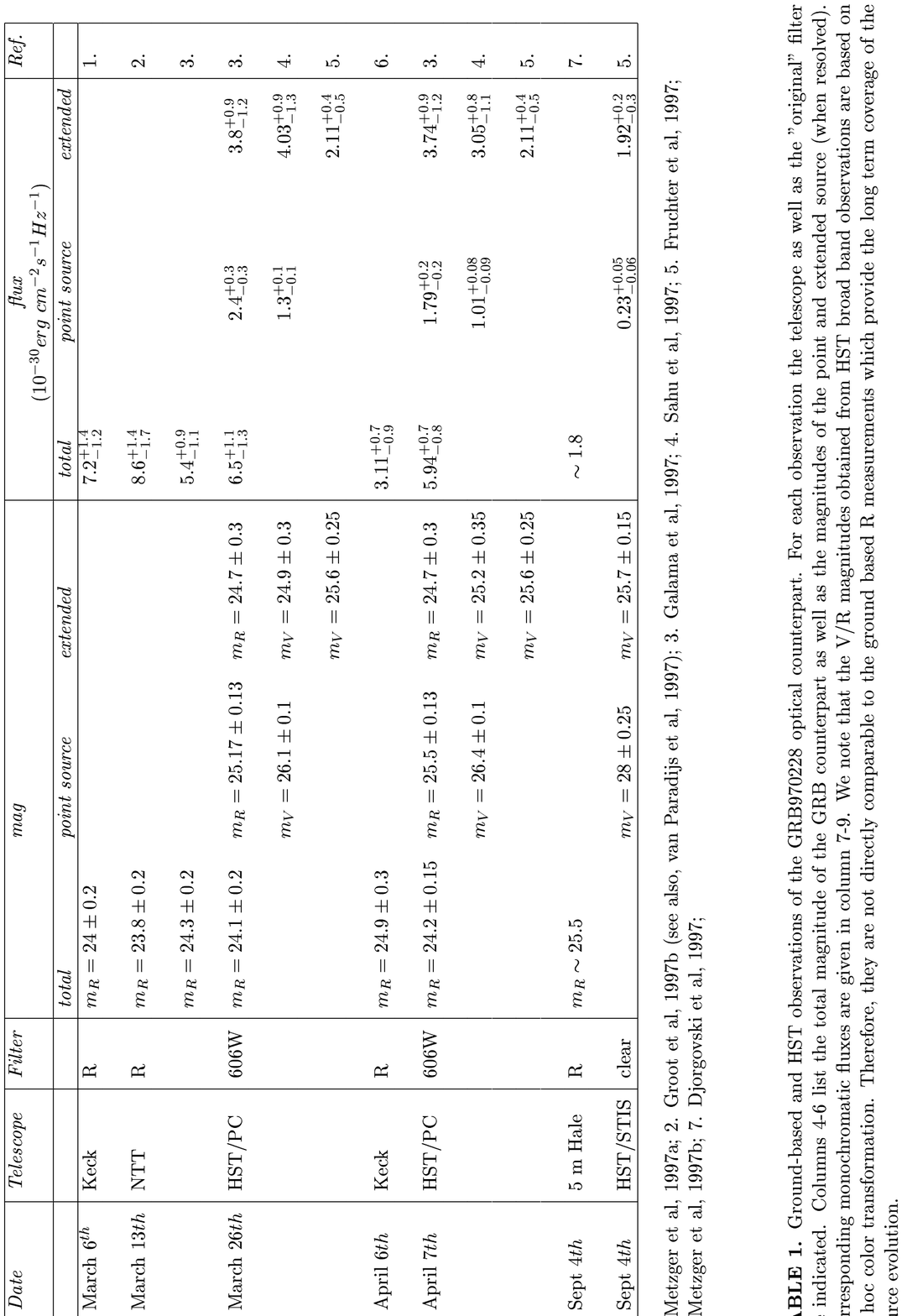,angle=180,height=23.5cm,width=17.5cm,clip=}}
 \refstepcounter{table} \label{sym_tab}
 \end{table}
 
Indeed, all the March ground  measurements agree in describing the  extended
object as elongated in the North-South direction with $m_{R} \sim 24$. HST
observations  were carried out in late March and early April using  the WFPC and
a broad V filter (F606). The extended source was resolved  into a point source
superimposed to a "fuzz", which, according to Sahu et  al. (1997), were detected
at  $m_{V} = 26.1 \pm 0.1$  and  $m_{V}=24.9 \pm 0.3$, respectively.  Comparison
of the March and April images showed that the  point source was most probably
fading, while nothing definite could be  said on the diffuse emission (Sahu et
al. 1997). Using the same data,  Galama et al (1997) estimate an R magnitude of
$m_{R}=25.17 \pm 0.13$ for the  point source and $m_{R} = 24.7 \pm 0.30$ for the
fuzz. In the same paper, the  value of the magnitude measured by the NTT on March
$13^{th}$ was also  revised, bringing it to $m_{R}=24.3 \pm 0.2$. More Keck observations
taken on April $5^{th}$ and $6^{th}$ gave, for the total  emission, a $m_{R}=24.9 \pm 0.3$ 
(Metzger et al. 1997b) i.e. significantly lower  than both the HST one and the
Keck March $6^{th}$ data. In April the source became unobservable from the ground and 
from HST. The observability window opened again in late August when  it was
pointed both from the Keck/Palomar (IAU Circ 6732) and from  HST using, this
time, the newly installed STIS. Both observations show the overall flux to be
lower than that measured  previsiously. On Sept $4^{th}$, Djorgovski et al (1997)
used the Palomar 5m  telescope to obtain an R image of the field where the
extended source was detected at $m_{R} \sim 25.5$. The STIS instrument on board
HST also  observed the source on Sept $4^{th}$ with the Clear filter.  Fruchter
et al,  (1997) are barely able to detect the point source, now at $m_{V}=28.0 
\pm 0.25$,  over a diffuse emission of $m_{V}=25.7 \pm 0.15$, i.e. 0.8 magnitude
fainter than in the HST March observation. This prompted a reanalysis of the 
March/April WFPC data which resulted in a reassessment of their  magnitude value
now estimated at $m_{V}=25.6 \pm 0.25$, i.e. half of the flux  published by Sahu
et al (1997) for two independent WFPC observations.   Although not stated, a
similar downward revision should apply also to the R  magnitude values published
by Galama et al (1997) for the HST  observations. \\ However, even accepting that
the HST data, after re-analysis, can be rendered consistent, it seems very
difficult to reconcile  the September STIS/Palomar data with the NTT/Keck ones of
early March. Table 1 shows  that a suggestion for fading of the extended
component of the OT seems to be present. However, magnitude  values do not always
render easy the comparison of data taken through  different filters. The
suggestion for a significant fading of the extended  source, implicit when
comparing March to September data, becomes  stronger when one computes the actual
energy fluxes. This is done in Table 1, where we have transformed the magnitude 
values in $erg/cm^{2} sec Hz$. 

\section*{HST vs NTT}

In the following, we shall compare the NTT data (kindly provided to us by Jan van
Paradijs) with the HST ones.  In order to  do so we have to assume that the
extended emission seen from the  ground is indeed the superposition of the
fuzziness seen by HST plus a  point source.  If we assume that the September 
STIS/Palomar
flux values for the extended source are correct, and if we further assume no
fading, we have to explain the extended total emission seen both by NTT and by
Keck with a combination of the STIS/Palomar fluxes plus a point source  of
suitable magnitude.  Even considering the revised NTT mag value given by Galama
et al  (1997), we have to account for a total flux of  $\sim 5.4~10^{-30}
erg/cm^{2}~sec ~Hz$.   Since the extended source observed in September  provides
$\sim 1.9~10^{-30} erg/cm^{2}~sec~Hz$, the  unseen point source should have been
$\sim 3.5~10^{-30} erg/cm^{2}~sec~Hz$, i.e.  definitely brighter that the
extended one.  However, in order to simulate the appearance of such a combination
one  should be able to locate the HST point source into the NTT nebulosity. This
calls for an accurate superposition of the HST March data onto the NTT/SUSI
frame.  To take care  of geometric distorsion, we have used the task "mosaic"
which re-scales  the HST/PC image, rebinning it to pixel size of 0.1". The
resulting image  has been rebinned and rotated onto the SUSI one (0.13"/pixel)
using a standard technique which is certainly accurate to better than  1/2 pixel
(actually 1/10 would be a more realistic estimate).  Figure 1  shows the
superposition of the HST March frame onto the SUSI one. Only the central  portion
of the actual images is given. Zooming on the OT,  one sees clearly that the HST
point source falls in the central part  of the NTT nebulosity, where the emission
is less intense and no hint of  a point-like object is seen.  However, the
central region of the nebulosity is just where one should put a hypothetical
point source of $3.5~10^{-30} erg/cm^{2}~sec~Hz$, corresponding to $m_{R} \sim
24.65$. This value is similar to the flux measured by the NTT for the faint
source just  south of the GRB970228 counterpart. Such a source is in interesting
test  case, since it is point-like in the HST/PC image (star \#3 in Figure 1) but
it looks extended in the NTT image.  However, inspection of Figure 2, where we
have compared the Right Ascension and Declination tracings of the two sources, 
shows unambigously their difference in shape for a comparable flux.   While star
\#3 is dominated  by a clear peak superimposed to a  region of higher background,
the GRB nebulosity does not show any  obvious point-like contribution.    This is
somewhat surprising, since a point source of $m_{R} \sim 24.6$ should have been
far easier to  detect than a $m_{R} \sim 25.5$ extended one. Moreover, we note
that such a faint extended source would be hardly within reach of an 1 hour NTT
exposure.  \\ Thus, the truly extended nature of the NTT source, coupled with the
lack of point source at the HST location, leads to the conclusion that the
nebulosity itself has  faded away from March $13^{th}$ and Sept $4^{th}$.

\begin{figure}[h!] 


\centerline{\hbox{
\psfig{file=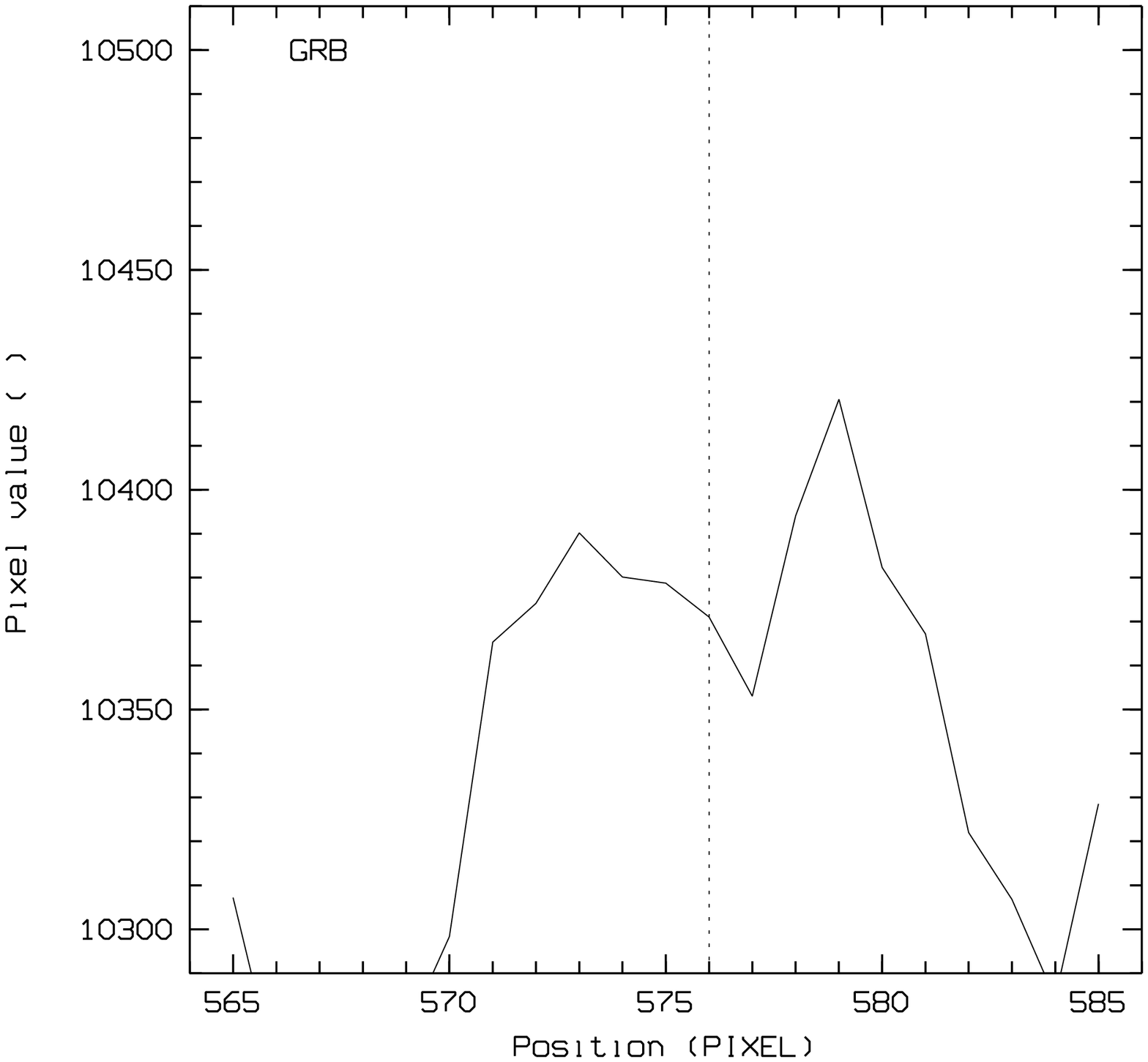,height=5cm,angle=270,clip=}
\psfig{file=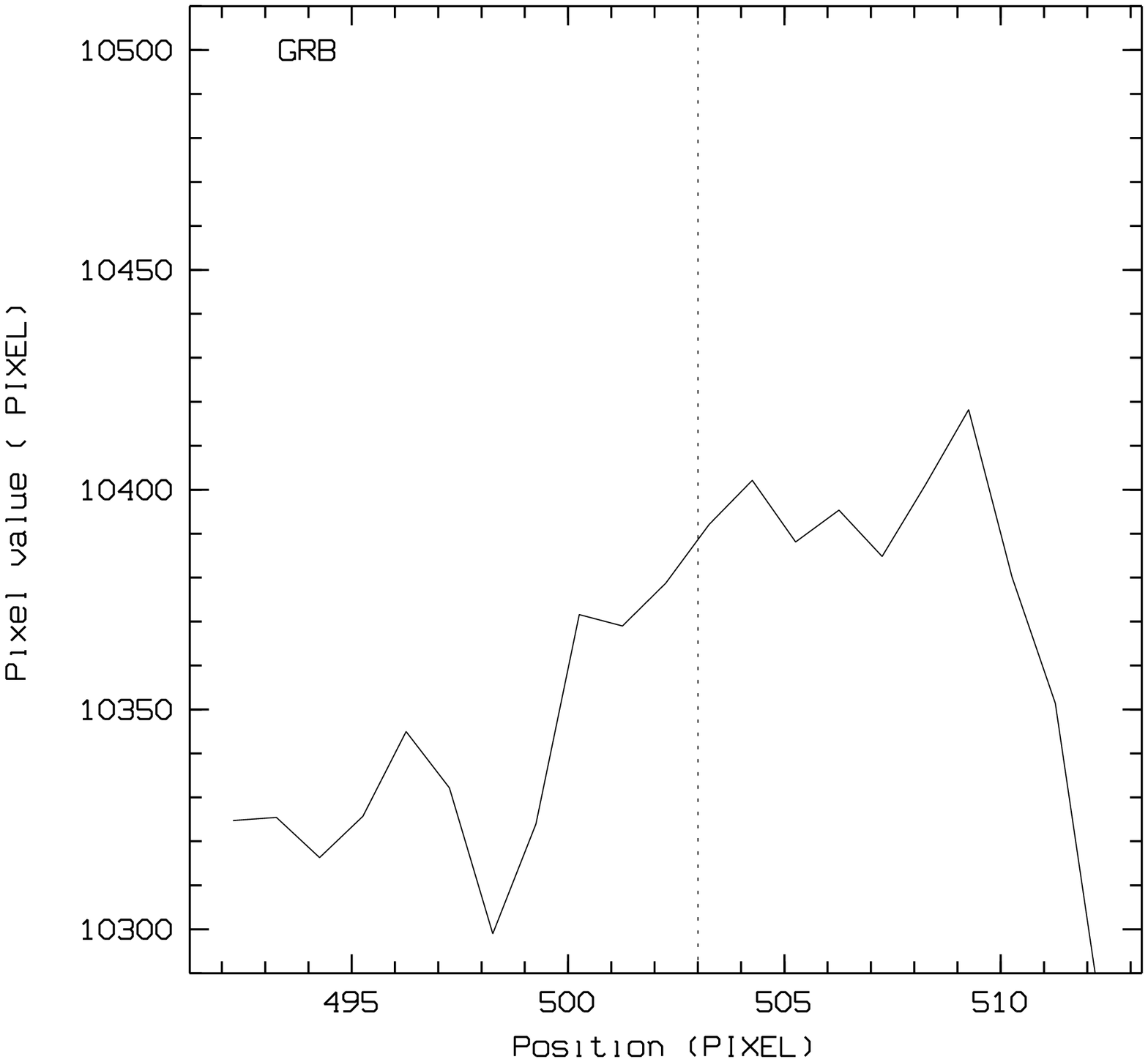,height=5cm,angle=270,clip=}
}}


\centerline{\hbox{
\psfig{file=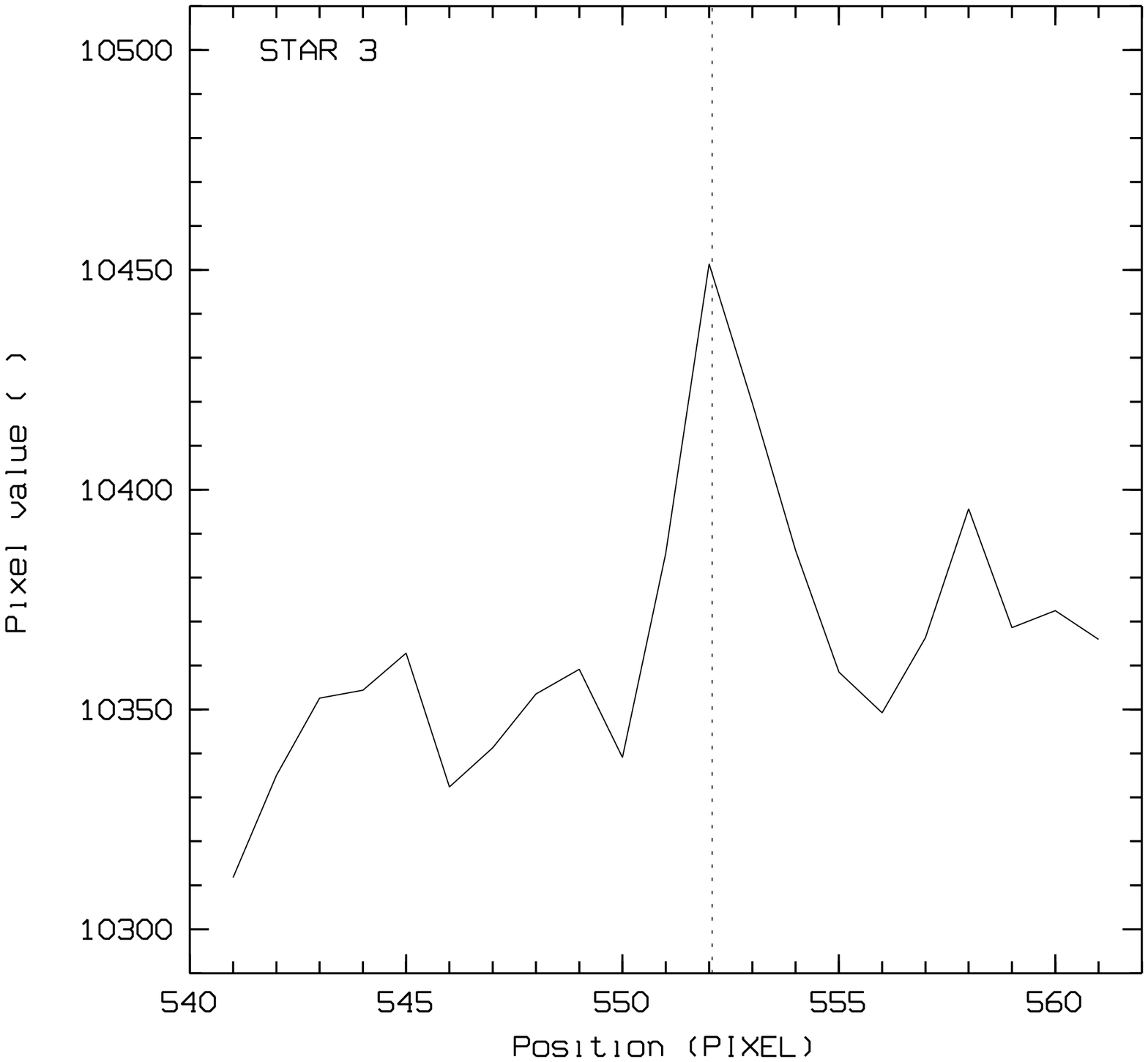,height=5cm,angle=270,clip=}
\psfig{file=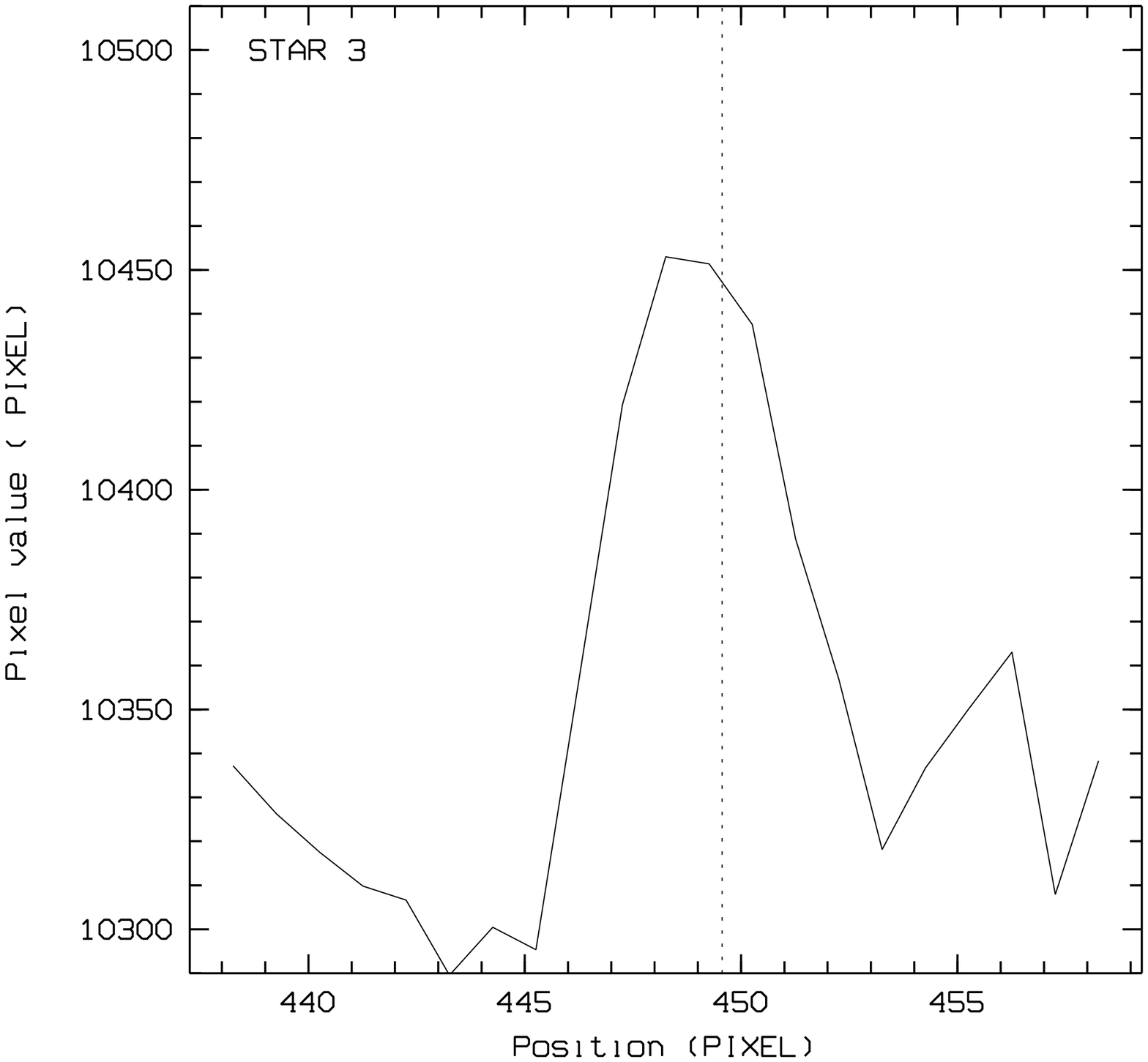,height=5cm,angle=270,clip=}
}}
\caption{Right ascension (left) and declination (right) tracings for the GRB
counterpart  (upper panels)  and for star \#3 (lower panels) obtained from the
NTT image shown in Fig.1.  In both cases, the tracings have been computed at the
expected location of the point sources. Their coordinates, obtained by registering
the  HST/PC frame onto the NTT/SUSI one, are
marked as vertical lines.
Although the PSF of star \#3 is affected by non-optimal seeing conditions ($\ge
1$ arcsec), which result in a profile broadening along declination, the presence
of  a point source is certainly recognazable.}
\label{}
\end{figure}

\section*{Conclusions}
Although comparing fluxes obtained with different instruments,  different filters
and different observing conditions is not  straightforward, the compilation of
the magnitudes values measured so  far for the optical counterpart of GRB970228
points toward a fading  both of the point source and of the diffuse emission.
While the fading of  the point source is expected in all theoretical scenarios,
the fading of the  diffuse emission has far reaching consequences and, as such,
is in need  of a dedicated observing campaign. The data available are numerous,
but  too diverse to provide the constraints needed to assess with certainty if, 
and how much, the nebulosity has faded. Indeed, for GRB 970228, it looks as if
every new observation results in a  downward revision of the values previously
published. Only more observations, directly comparable with those already in 
hand (i.e. obtained with identical instrumental-ups)  can provide a  definite
aswer to this all important point. Of particular importance could be new HST/PC
data since the unfiltered STIS  image is not directly comparable to the PC ones, 
obtained with a broad V filter.


\begin{references}
\bibitem{}Caraveo P.A. et al,{\it A \& A.}\ {\bf 326}, L13 (1997).
\bibitem{}Costa E. et al.{\it Nature} {\bf 387}, 783 (1997).
\bibitem{}Djorgovski S. et al {\it IAU Circ 6732} (1997)
\bibitem{}Fruchter, A. et al. {\it IAU Circ. 6747}(1997)
\bibitem{}Galama T. et al. {\it Nature} {\bf 387} 479 (1997)
\bibitem{}Groot J.P. et al. {\it IAU Circ. 6584}  (1997a)
\bibitem{}Groot J.P. et al. {\it IAU Circ. 6588}  (1997b)
\bibitem{}Metzger M.R. et al. {\it IAU Circ. 6588}  (1997a)
\bibitem{}Metzger M.R et al. {\it IAU Circ. 6631}  (1997b)
\bibitem{}Sahu K. et al. {\it Nature} {\bf 387} 476 (1997)
\bibitem{}van Paradijs et al.{\it Nature} {\bf 386} 686 (1997)
\end{references}
\end{document}